*Preprint, November 2026. This preprint has not been peer-reviewed*

# High resolution 3D localization of ultrasound sparse sources using a SAGE algorithm

Sophie V. Heymans[1,2], François Rottenberg[3], Tatiana Gelvez-Barrera[4,5], Adrian Basarab[4,*] and Barbara Nicolas[4,*]

[1]Department of Physics, KU Leuven campus KULAK, 8500 Kortrijk, Belgium
[2]Department of Cardiovascular Sciences, KU Leuven, 3000 Leuven, Belgium
[3]Department of Electrical Engineering, KU Leuven, 9000 Gent, Belgium
[4]CREATIS, UMR5220, CNRS, Université Claude Bernard Lyon 1, INSA-Lyon, Inserm U1294, 69100 Lyon, France
[5]IRIT, Université de Toulouse, CNRS, 31062 Toulouse, France

Email: sophie.heymans@kuleuven.be

*This work was supported by the Research Foundation-Flanders (FWO) under the Junior Postdoctoral Fellowship grant number 12E0923N, and by the LABEX CELYA (ANR-10- LABX-0060) of Université de Lyon, within the program "Investissements d'Avenir" (ANR-11-IDEX-0007) operated by the French National Research Agency (ANR).*

[*]*A. Basarab and B. Nicolas contributed equally to this work*

## Abstract

Reliable localization and characterization of cavitation activity are central to many ultrasound imaging and therapy monitoring applications. Passive acoustic mapping (PAM) can estimate cavitation distributions without prior assumptions but remains limited by the system point spread function. While advanced PAM methods improve spatial resolution, their performance may degrade for transient emissions with uncertain emission times. In this work, we propose a parametric source localization framework based on the Space-Alternating Generalized Expectation-Maximization (SAGE) algorithm to estimate source positions and emitted waveforms directly from array channel data. A greedy source extraction and localization procedure is first used to initialize the model order and source parameters, after which SAGE iteratively refines each source estimate. The proposed method was validated on 49 simulated acquisitions containing up to seven simultaneously emitting sources with emission time uncertainties of 12 µs, achieving a source localization precision of 0.24 mm (standard deviation of the total localization error). Experimental evaluation on proton-induced nanodroplet vaporization data acquired with an ultrasound matrix array demonstrated improved spatial resolution compared to a beamforming-based approach and sub-millimeter agreement between vaporization maps and the proton distribution in all three spatial dimensions. These results highlight the potential of parametric source localization for high-resolution proton therapy monitoring and provide a general framework for the localization of simultaneously emitting sparse ultrasound sources.

## I. Introduction

Accurate localization of acoustic emissions from ultrasound contrast agents (UCAs) and cavitation nuclei is central to a growing range of ultrasound imaging and therapy monitoring applications. In imaging, Ultrasound Localization Microscopy (ULM) exploits the sparse distribution of microbubbles to visualize the microvasculature at resolution beyond the diffraction limit [1], [2]. In therapy monitoring, Passive Acoustic Mapping (PAM) enables to characterize cavitation activity during interventions including blood brain barrier opening, ultrasound-mediated drug and gene delivery, histotripsy, or sonothrombolysis, [3], [4]. Acoustic cavitation induced by non-ultrasonic energy sources can also be exploited for therapy monitoring, for example in laser lithotripsy [5] or in radiosensitive nanodroplet (ND) vaporization detection for radiotherapy dosimetry or proton range verification [6], [7], [8].

Acoustic source localization with ultrasound arrays is generally referred to as beamforming. In both imaging and therapy monitoring applications, a high spatial resolution is essential to ensure sufficient clinical performances in terms of accuracy, safety and efficacy [9], [10]. However, the acoustic emissions encountered in these applications span a wide range of temporal behaviors, from quasi-stationary oscillations to short transient events [11], [12], making accurate and high-resolution source localization particularly challenging.

Beamforming can be performed both in the time-domain (TD) or in the frequency-domain (FD). The most fundamental TD beamforming approach, delay-and-sum (DAS) [13], is based on the coherent summation of signals delayed by the theoretical travel time between the transducer elements and the candidate source location, assuming a known speed of sound. When the activation time of the source is known, a Time of Arrival (TOA) strategy can be used with delays combining the travel time from emitter to target and from target to receivers. In this scenario, the spatial resolution is determined primarily by the duration of the received echo [14]. This approach is used in active imaging with short pulses, but also in therapy using short pulses and temporal synchronization between the therapy and imaging system, *i.e.* ATP-PAM [15] or power cavitation imaging (PCI) [14], [16]. In many situations, however, cavitation sources are exposed to long (quasi-continuous) excitations and may emit at any time during the exposure.

Therefore, time exposure acoustics (TEA), which relies on the Time Differences of Arrival (TDoA) between the source location and the array elements, is employed [17], [18], The cavitation maps are integrated over the exposure duration (Delay, sum and integrate or DSAI) and the spatial resolution is governed by the diffraction pattern of the receive array [14]. Adaptive versions of the DSAI, such as the Robust Capon beamformer (RCB-PAM) or approaches weighting DSAI with the phase or amplitude coherence factor [4], [19], have been proposed to improve the spatial resolution, at the cost of computational complexity [20]. In addition, when assumptions about the sources can be made, sparse reconstruction/model-based approaches can be used to increase the image resolution. For instance, the temporal sparsity of inertial cavitation events was exploited through the use of sparse deconvolution techniques, namely matching pursuit and basis pursuit, in [21]; later, PAM based on compressed sensing was proposed [22], [23]. A model-based TD-inverse problem was also recently introduced to retrieve spatiotemporal cavitation maps under several regularization strategies [24].

As an alternative to TD beamforming, FD beamforming has gained popularity over the years, as frequency selection in the Fourier domain allows both to strongly reduce the computation time and to characterize the cavitation regime based on the emissions spectral content. The DSAI approach can be readily translated to the FD by applying phase shifts to the Fourier transform of received signals instead of temporal delays [25]. FD adaptive beamforming methods, *e.g.* the Pisarenko and MUSIC beamformers [26], and inverse methods based on the Cross Spectral Matrix (CSM) fitting [27], have also been introduced, yielding major improvements in image resolution.

While PAM has undergone significant advances, achieving high spatial resolution while maintaining robustness to transient emissions remains a fundamental challenge. Advanced FD methods can achieve super-resolution, but rely on accurate estimation of the CSM and therefore require sufficiently long and quasi-stationary acoustic emissions and averaging over multiple temporal windows [26]. As a result, they are poorly suited to short-lived events such as ND vaporization or inertial cavitation bursts. Conversely, TD approaches remain applicable to transient emissions, but their localization performance is often limited by timing uncertainties arising from finite excitation pulse lengths, synchronization errors, and unknown source activation or emission durations. TOA-based methods such as PCI and ATP-PAM assume precise knowledge of the excitation timing and implicitly link the source emission duration to the excitation duration [8], [9]. Consequently, their axial resolution deteriorates rapidly as the pulse length increases, ultimately converging to TEA-PAM [8] and reaching the centimeter range in practical situations. To the best of our knowledge, no PAM approach has demonstrated simultaneous high-resolution localization of sparse transient cavitation events while remaining robust to substantial timing uncertainties. Such capability is particularly important for monitoring short-lived phenomena including ND activation or inertial bubble collapse, for which accurate spatiotemporal characterization is critical to warrant treatment safety and efficacy.

One possible route to overcome this trade-off is provided by parametric source localization methods. Rather than reconstructing a source distribution on a predefined spatial grid, these approaches rely on a parametric model of the received signals and directly estimate a limited set of source parameters from the measurements. Their main advantage is the ability to achieve a localization accuracy beyond the diffraction limit and detailed source characterization from a relatively small amount of data. However, their performance strongly depends on the validity of the assumed signal model and on prior knowledge regarding the number and nature of the sources. Within this framework, maximum-likelihood estimators were proposed for the analysis of ultrasonic backscattered echoes [28]. The measured signal was modeled as a superposition of echoes originating from multiple scatterers, each characterized by parameters such as arrival time, amplitude, phase, center frequency, and bandwidth. As direct joint estimation yields a highly nonlinear, high-dimensional optimization problem, Expectation-Maximization (EM) and Space-Alternating Generalized Expectation-Maximization (SAGE) algorithms estimate source parameters sequentially, thereby reducing problem dimensionality while retaining the maximum-likelihood formulation [28], [29]. While such approaches have been successfully applied in quantitative ultrasound for bone characterization [30], [31], their application to passive ultrasound source localization has, to the best of our knowledge, not yet been reported.

In this contribution, we introduce a parametric approach based on the SAGE algorithm, which combines the super-resolution potential of parametric estimation with the flexibility required for transient cavitation emissions. Specifically, it aims at recovering the source positions and emitted waveforms without requiring prior knowledge of the emission timing and duration (unlike in TOA-based beamforming), or spectral content (unlike in FD approaches). When partial timing information is available, it can be incorporated as prior information to further improve the localization accuracy. Our approach was evaluated using simulated and experimental data of radiation-induced ND vaporization. The proposed SAGE framework was benchmarked against two alternative localization strategies: conventional TD-DAS beamforming followed by source super-localization, and a greedy source extraction (GSE) method inspired by matching pursuit, which also provides the parameter initialization for SAGE.

The remainder of this paper is organized as follows. Section II introduces the source model and the proposed GSE and SAGE methods. The simulated and experimental datasets are described in Section III, and the results are presented in Section IV and discussed in Section V.

## II. Proposed localization methodologies

### A. General problem description and system model

An ultrasound (matrix) array with $M$ elements located at positions $\boldsymbol{r_m} = (x_m, y_m, 0)$ is receiving acoustic signals from an unknown number ($L$) of sources (with $L < M$), assumed to be small such that it results in sparse beamformed maps. The sources positions are assumed to remain stationary at the time scale of the radiofrequency (RF) data acquisition (<1 ms). Each source has a different emission start time $t_{0,l}$ and 3D position $\boldsymbol{r}_{S,l} = (x_{S,l}, y_{S,l}, z_{S,l})$. We assume an homogeneous, non-attenuating medium with constant speed of sound $c$. If each source $l$, located at position $\boldsymbol{r}_{S,l}$, emits a signal $s_l(t)$, the RF signal recorded by array element $m$ can be modelled as:

$$y_m(t) = \sum_{l=1}^{L} h_m(\boldsymbol{r}_{S,l}, t) * s_l\left(t - \tau_m(\boldsymbol{r}_{S,l})\right) + \eta_m(t) \qquad (1)$$

In this expression, $\eta_m(t)$ is additive noise, assumed to be white Gaussian, $h_m(\boldsymbol{r}_{S,l}, t)$ is the known spatial impulse response of element $m$ to a source located at position $\boldsymbol{r}_{S,l}$, $*$ is the convolution product and $\tau_m(\boldsymbol{r}_{S,l})$ is the propagation delay from source $l$ to element $m$, given by:

$$\tau_m(\boldsymbol{r}_{S,l}) = t_{0,l} + \frac{\|\boldsymbol{r}_{S,l} - \boldsymbol{r}_m\|}{c}. \qquad (2)$$

The recorded RF data are obtained by sampling $y_m(t)$ at $t = nT_s$, with $n = 0,1,\dots,N_t$ time samples and $T_s$ the sampling period, yielding $\boldsymbol{y}_m[n] = y_m(nT_s)$, with $\boldsymbol{y}_m \in \mathbb{R}^{N_t}$ the RF data vector of array element $m$.

## B. TD Delay-and-sum beamforming approach

The objective of beamforming is to estimate a datacube representing the total cavitation activity across space and time. The beamformed map $b$ should ideally estimate the signal emitted by each source, $s_l(t)$, at its corresponding location $\boldsymbol{r}_{S,l}$, and consist of zeros elsewhere, *i.e.,* if δ denotes a Dirac delta function:

$$b(\boldsymbol{r}, t) = \sum_{l=1}^{L} s_l(t)\delta(\boldsymbol{r} - \boldsymbol{r}_{S,l}).$$

Its most simple form, 4D TD-DAS, is achieved by delaying the channel data by the propagation times to each voxel:

$$\hat{b}(\boldsymbol{r}, t) = \sum_{m=1}^{M} w_m y_m(t + \tau_m(\boldsymbol{r})) \qquad (3)$$

in which $\tau_m(\boldsymbol{r})$ is the propagation delay from pixel $\boldsymbol{r}$ to element $m$, and $w_m$ are optional receive apodization weights. In practice, this consists in beamforming a 3D volume for every time point present in the RF data, yielding a datacube sampled at $N_x \times N_y \times N_z$ spatial voxels and $N_t$ temporal frames, *i.e.*, when spatial dimensions are stacked together: $\widehat{\boldsymbol{B}}[i, n] = \hat{b}(\boldsymbol{r}, t)$, with $i = 0,1,\dots,N_xN_yN_z$ and $\widehat{\boldsymbol{B}} \in \mathbb{R}^{N_xN_yN_z \times N_t}$. While TD-DAS beamforming estimates the cavitation activity distribution without prior assumptions, it does not directly recover the true source locations. Instead, the reconstructed map represents the source distribution convolved with the system point spread function (PSF), which is typically elongated in the axial direction in PAM due to the unknown source emission times ($t_{0,l}$).

## C. Proposed model-based approaches

Rather than estimating a cavitation datacube on a dense 4D grid, our proposed parametric approach seeks to directly estimate sparse source parameters: $L \times 3$ source coordinates $\boldsymbol{r}_{S,l}$ and $L \times N_t$ emitted signals $s_l(t)$, based on the RF data observations $y_m(t)$. The source positions allow to accurately localize cavitation, while emitted signals offer insights into the cavitation regime (and hence associated bioeffects). Each source is characterized by a parameter set $\boldsymbol{\theta}_l = \{\boldsymbol{r}_{S,l}, \boldsymbol{s}_l\}$ with $\boldsymbol{r}_{S,l} \in \mathbb{R}^3$ and $\boldsymbol{s}_l = [s_l(T_s)\ s_l(2T_s)\ \cdots\ s_l(nT_s)]^T \in \mathbb{R}^{N_t}$, $(.)^T$ denoting the transpose operation. The complete parameter set to retrieve is then $\boldsymbol{\Theta} = \{\boldsymbol{\theta}_1, \dots, \boldsymbol{\theta}_L\}$.

We first present a coarse grid-based iterative greedy approach for source extraction and localization, inspired by matching pursuit [30]. Next, we introduce a SAGE-based iterative procedure that sequentially refines source parameters. In the expectation or E-step, the contribution of a given source to the observed RF data is isolated by removing the currently estimated contributions of all other sources. The maximization or M-step then updates the source parameters via maximum-likelihood estimation. These successive source-wise updates are repeated until convergence. As the SAGE method is gridless, the achievable resolution is independent from a chosen voxel size.

### *1. Greedy source extraction and localization (GSE)*

The method follows a greedy detect-estimate-subtract strategy, whereby sources are recovered sequentially from the residual RF data. After localizing and estimating the strongest source, its modeled contribution is subtracted from the measurements, and the procedure is repeated until no additional sources are detected. When, contrarily to matching pursuit, no dictionary of hypothetical emitted signals can be derived, the emitted signals of each source are estimated by beamforming. The steps are detailed below and summarized in Algorithm 1.

First, the source index, $l$, is set to 1 and the estimated residual RF data, $\hat{y}_{m,\mathrm{res}}^{(0)}(t)$, is initialized as the observation $y_m(t)$. The superscript $(.)^{(l)}$ refers to the source index. The 4D datacube $\hat{b}(\boldsymbol{r}, t)$ is beamformed using (3) on the residual channel data and multiplied by the coherence factor (CF) defined in [4] to improve its resolution:

$$\hat{b}_{\mathrm{CF}}(\boldsymbol{r}, t) = \hat{b}(\boldsymbol{r}, t) \frac{\left|\sum_{m=1}^{M_{\mathrm{act}}} \hat{y}_{m,\mathrm{res}}^{(l-1)}(t + \tau_m(\boldsymbol{r}))\right|^2}{M_{\mathrm{act}} \sum_{m=1}^{M_{act}} \left|\hat{y}_{m,\mathrm{res}}^{(l-1)}(t + \tau_m(\boldsymbol{r}))\right|^2} \qquad (4)$$

In this equation, $M_{\mathrm{act}}$ is the number of elements used in the active aperture when receive apodization is applied. If prior knowledge of the source emission time range ($t_{0,l}$) is available, the 4D datacube can be bounded in time. This is often the case when cavitation is induced by a pulsed excitation source (ultrasound, radiation, laser, etc.) synchronized with data acquisition.

In this 4D datacube, the strongest source is assumed to lie in the largest cluster of data points with amplitude above a fixed threshold. This enables to select a subset of candidate voxels ($\boldsymbol{r}_{\mathbf{cand}}$) for source $l$. For those voxels, the known beamformed signals $\hat{b}(\boldsymbol{r}_{\mathbf{cand}}, t)$ obtained from (3) without CF weighting and with no bounds in the time dimension (to preserve the full duration of $s_l(t)$) are used to compute the expected RF data from a single source located at this voxel:

$$\hat{y}_{m,l}(\boldsymbol{r}_{\mathbf{cand}}, t) = h_m(\boldsymbol{r}_{\mathbf{cand}}, t) * \hat{b}(\boldsymbol{r}_{\mathbf{cand}}, t - \tau_m(\boldsymbol{r}_{\mathbf{cand}})). \qquad (5)$$

The source position is then determined as the voxel minimizing the sum of squared errors (SSE) between the residual channel data and the source channel data from (5).

$$\hat{\boldsymbol{r}}_{S,l} \;=\; \arg\min_{\boldsymbol{r}_{\mathbf{cand}}} \sum_{m=1}^{M}\sum_{t=1}^{N_t}\left(\hat{y}_{m,\mathrm{res}}^{(l-1)}(t) - \hat{y}_{m,l}\left(\boldsymbol{r}_{\mathbf{cand}}, t\right)\right)^2. \quad (6)$$

The source cavitation signal is then obtained as:

$$\hat{s}_l(t) \;= b(\hat{\boldsymbol{r}}_{S,l}, t). \quad (7)$$

The estimated RF data generated by the source is given by substituting $\boldsymbol{r}_{\mathbf{cand}}$ by $\hat{\boldsymbol{r}}_{S,l}$ in (5):

$$\hat{y}_{m,l}(t) = \hat{y}_m\left(\hat{\boldsymbol{r}}_{S,l}, t\right). \quad (8)$$

Then, the residuals are updated by subtracting the contribution from the localized source:

$$\hat{y}_{m,\mathrm{res}}^{(l)}(t) = \hat{y}_{m,\mathrm{res}}^{(l-1)}(t) - \hat{y}_{m,l}(t), \quad (9)$$

and the process is repeated by incrementing the source index $l$. The iterations stop once the maximum number of sources $L_{\max}$ has been initialized, the residual energy of the channel data falls below a threshold (computed based on acquisitions in the absence of sources) or the decrease in channel data energy due to the subtraction of source $l$ is inferior to an user-defined threshold.

Owing to array diffraction and errors in the estimated source position and emitted signal, complete source cancellation is not always possible. The remaining residuals can subsequently trigger additional detections of the same source. Therefore, sources located within a predefined distance threshold of one another were merged (Algorithm 1, line 20): for each cluster of neighboring sources, the estimated source signals $\hat{s}_l(t)$ and RF data contributions $\hat{y}_{m,l}(t)$ were added together and represented as a single source. The resulting source location was taken as that of the strongest source in the cluster.

**Algorithm 1 – GSE**

**Require:** $y_m(t), h_m(\boldsymbol{r}, t), L_{\max}$, energy decrease threshold $\varepsilon$, residual energy threshold $E_{\mathrm{th}}$, cluster selection threshold $I_{\mathrm{th}}$

1: Initialize: $l = 1, \hat{y}_{m,\mathrm{res}}^{(0)}(t) = y_m(t)$
2: **While** $l \le L_{max}$ **do**
3: $\hat{b}(\boldsymbol{r}, t)$ via (3) applied to $\hat{y}_{m,\mathrm{res}}^{(l-1)}(t)$
4: $\hat{b}_{\mathrm{CF}}(\boldsymbol{r}, t)$ via (4)
5: $\boldsymbol{r}_{\mathbf{cand}} \leftarrow$ LargestCluster$\{\boldsymbol{r} : \hat{b}_{\mathrm{CF}} > I_{th}\}$
6: **For** each $\boldsymbol{r}_{\mathbf{cand}}$ **do**
7: $\hat{y}_{m,l}(\boldsymbol{r}_{\mathbf{cand}}, t)$ via (5)
8: $j(\boldsymbol{r}_{\mathbf{cand}}) = \sum_m \sum_t \left(\hat{y}_{m,\mathrm{res}}^{(l-1)}(t) - \hat{y}_{m,l}(\boldsymbol{r}_{\mathbf{cand}}, t)\right)^2$
9: **End for**
10: $\hat{\boldsymbol{r}}_{S,l} = \arg\min_{\boldsymbol{r}_{\mathbf{cand}}} j(\boldsymbol{r}_{\mathbf{cand}})$ via (6)
11: $\hat{s}_l(t)$ via (7)
12: $\hat{y}_{m,l}(t)$ via (8)
13: $\hat{y}_{m,\mathrm{res}}^{(l)}(t)$ via (9)
14: **If** $\left|\sum_m \sum_t \hat{y}_{m,\mathrm{res}}^{2\,(l)}(t) - \sum_m \sum_t \hat{y}_{m,\mathrm{res}}^{2\,(l-1)}(t)\right| < \varepsilon$ or $\sum_m \sum_t \hat{y}_{m,\mathrm{res}}^{2\,(l)}(t) < E_{\mathrm{th}}$ **do**
15: **exit**
16: **else**
17: $l = l + 1$
18: **End if**
19: **End while**
20: Neighbour merging
21: **Return** $L, \boldsymbol{\Theta}, \hat{y}_{m,l}(t)$

*2. Source parameter estimation via SAGE*

The GSE procedure provides an initial estimate of the source position and waveforms. To further refine these estimates, we employ a SAGE-based iterative procedure. In each iteration, the contribution of a single source is isolated from the observed RF data (E-step) and its position is updated while keeping all other sources fixed (M-step). Therefore, $L$ successive EM steps need to be performed to optimize all source parameters and complete a full iteration cycle.

The proposed implementation differs from conventional SAGE by updating the $N_t$ parameters of the source waveform $s_l(t)$ separately through beamforming rather than jointly with the source position in the M-step. Therefore, the parameter set to optimize via the SAGE loop was $\boldsymbol{\theta'}_l = \{\boldsymbol{r}_{S,l}\}$. This decoupling of source position and waveform estimation improves robustness to imperfect source cancellation (due to model mismatch commonly encountered in experimental data, inhomogeneities in individual elements sensitivity, and medium effects such as attenuation, speed of sound variations and scattering), thereby stabilizing the recovery of weak sources. The proposed SAGE variant is detailed below and in Algorithm 2, while a derivation of the conventional formulation, jointly optimizing the source positions and emitted signals, is provided in the appendix.

The GSE method presented above was used to initialize the source parameters as $\boldsymbol{\Theta}^{(0)} = \boldsymbol{\Theta}_{\mathrm{GSE}}$, with the superscript $(.)^{(0)}$ referring to the SAGE initialization. In the E-step, the unobserved RF data generated by source $l$ is estimated by subtracting from the observed RF data the predicted RF data for all sources except the one under consideration, *e.g.*, for iteration step $k$ [32]:

$$\hat{y}_{m,l}^{(k)}(t) = \; y_m(t) - \sum_{l'=1}^{l-1} \hat{y}_{m,l'}^{(k)}(t) - \sum_{l'=l+1}^{L} \hat{y}_{m,l'}^{(k-1)}(t) \quad (10)$$

Next, the M-step estimates the parameters of source $l$ by minimizing the difference between the expectation (10) and the parametric model of the source (1):

$$\hat{\boldsymbol{\theta}}_l^{(k)} \;=\; \arg\min_{\boldsymbol{\theta}} \sum_{m=1}^{M}\sum_{t=1}^{N_t}\Big(\hat{y}_{m,l}^{(k)}(t) - h_m(\boldsymbol{r}, t) * s_l(t - \tau_m(\boldsymbol{r}))\Big)^2 \quad (11)$$

To keep the source signal fixed in the M-step, the source signal from the previous iteration $\hat{s}_l^{(k-1)}$ is first shifted by $\Delta\tau(\boldsymbol{r})$ to account for the path difference between the position being evaluated, $\boldsymbol{r}$, and the position at which the source signal model $\hat{s}_l^{(k-1)}$ was estimated, $\hat{\boldsymbol{r}}_{S,l}^{(k-1)}$:

$$\Delta\tau(\boldsymbol{r}) = \frac{1}{cM}\sum_{m=1}^{M}\left(\left\|\hat{\boldsymbol{r}}_{S,l}^{(k-1)} - \boldsymbol{r}_m\right\| - \left\|\boldsymbol{r} - \boldsymbol{r}_m\right\|\right) \quad (12)$$

This shifted source signal is back-propagated to the array to yield the modelled RF data to compare to the source RF data, in the following modified M-step:

$$\hat{\boldsymbol{\theta}}'^{(k)}_{l} = \arg\min_{\boldsymbol{\theta}'} \sum_{m=1}^{M}\sum_{t=1}^{N_t}\Big(\hat{y}^{(k)}_{m,l}(t) - h_m(\boldsymbol{r},t) * \hat{s}^{(k-1)}_{l}(t + \Delta\tau(\boldsymbol{r}) - \tau_m(\boldsymbol{r}))\Big)^2 \quad (13)$$

After solving (13) for $\boldsymbol{r}$, the source signal and estimated RF data for source $l$ are updated:

$$\hat{s}^{(k)}_{l}(t) = \hat{s}^{(k-1)}_{l}\left(t + \Delta\tau(\hat{\boldsymbol{r}}^{(k)}_{\boldsymbol{S},l})\right) \quad (14)$$

$$\hat{y}^{(k)}_{m,l}(t) = \hat{a}^{(k)}_{m,l}\left[h_m\left(\hat{\boldsymbol{r}}^{(k)}_{\boldsymbol{S},l}, t\right) * \hat{s}^{(k)}_{l}\left(t - \tau_m(\hat{\boldsymbol{r}}^{(k)}_{\boldsymbol{S},l})\right)\right] \quad (15)$$

In (15), we included element-specific non-negative normalization factors $a_{m,l}$ to account for experimental differences in element sensitivity. Those were obtained by finding the least-squares projection of the modelled RF data onto the (estimated) observed RF data for each source.

Once all $L$ sources have been updated, their emitted signals are further updated through beamforming using

$$\hat{s}^{(k)}_{l}(t) = \sum_{m=1}^{M} w_m \hat{y}^{(k)}_{m,l}\left(t + \tau_m(\hat{\boldsymbol{r}}^{(k)}_{\boldsymbol{S},l})\right) \quad (16)$$

and the estimated RF data are updated accordingly using (15). This additional step was included to ensure that the source waveform could vary, as (14) only performs a time shift. Equation (16) could have replaced (14) as well, but we noticed that this affected convergence; therefore, we chose to only include a time shift in the EM routine and to update the source waveforms only after all source positions have been updated.

The algorithm exits when the parameter distances between iteration $k-1$ and $k$, $d_\theta$, is below a threshold, or if the maximum number of iterations is reached. Due to the non-parametrization of the source signal and its update in (16), convergence of the algorithm is no longer guaranteed. Therefore, an exit criterion was introduced to detect divergence when the relative increase in total SSE exceeds a predefined threshold. In such cases, the iteration step $k$ corresponding to the minimum SSE is retained as the final result. In addition, sources are discarded when the CF, coefficient of determination, and RF data energy reduction fall below predefined thresholds (source pruning, line 12).

**Algorithm 2 – Proposed SAGE variant**

**Require:** $y_m(t), h_m(\boldsymbol{r},t), L, \boldsymbol{\Theta}^{(0)}, \hat{y}^{(0)}_{m,l}(t), k_{\max}$, parameter threshold $d_\theta$, SSE divergence threshold $d_{\text{SSE}}$

1: Initialize: $k = 1$
2: **While** $k \le k_{\max}$ **do**
3: **For** $l = 1, \dots, L$ **do**
4: $\hat{y}^{(k)}_{m,l}(t)$ via (10) — **E-step**
5: $\hat{\boldsymbol{\theta}}'^{(k)}_{l}$ via (13), using $\Delta\tau(\boldsymbol{r})$ from (12) — **M-step**
6: $\hat{s}^{(k)}_{l}(t)$ via (14): time-shift
7: $\hat{y}^{(k)}_{m,l}(t)$ via (15)
8: **End for**
9: Neighbour merging
10: $\hat{s}^{(k)}_{l}(t)$ via (16): beamforming
11: $\hat{y}^{(k)}_{m,l}(t)$ via (15)
12: Source pruning
13: $SSE^{(k)}_{\text{tot}} = \sum_m \sum_t \left(y_m(t) - \sum_l \hat{y}^{(k)}_{m,l}\right)^2$
14: **If** $\max_l \lVert \hat{\boldsymbol{r}}^{(k)}_{\boldsymbol{S},l} - \hat{\boldsymbol{r}}^{(k-1)}_{\boldsymbol{S},l} \rVert < d_\theta$ **do**
15: **exit**
16: **Else if** $SSE^{(k)}_{\text{tot}} / SSE^{(k-1)}_{\text{tot}} > d_{\text{SSE}}$
17: $\boldsymbol{\Theta}^{(k)} = \boldsymbol{\Theta}^{(\text{best})}$
18: **exit**
19: **Else**
20: $k = k + 1$
21: **End if**
22: **End while**
23: **Return** $L, \boldsymbol{\Theta}^{(k)}, \hat{y}^{(k)}_{m,l}(t)$

### 3. *Estimation of source emission start time*

The emission start times $t_{0,l}$ (see (2)) were not included in the optimization parameters as estimation of the source signals by beamforming across space and time allowed to preserve this temporal information, which is intrinsically present in the observed RF data. Therefore, the emission start time of each source was retrieved by finding the time index of the peak of the envelope of the estimated source signal $\hat{s}_l(t)$.

# III. Materials and Methods

## A. Simulations

Ultrasound RF data of sparse ND vaporization signals were simulated as follows. ND vaporization signals $s_l(t)$ were sampled from previous experiments of radiation-induced vaporization of superheated decafluorobutane (PFB) NDs with a polyvinyl alcohol shell dispersed in water [33]. The simulated matrix array was an 11 × 11 2 MHz large element array (element size 2.56 mm$^2$, aperture size 30.9 mm$^2$, Fraunhofer IBMT, Germany) used in those experiments. For each source, we randomly selected the emission start time $t_{0,l}$ in an interval from 0 to 8 µs and the source position $\boldsymbol{r}_{\boldsymbol{S},l}$ within a cubic volume of 26 mm (azimuth - *x*) by 26 mm (elevation - *y*) by 48 mm (axial - *z*). RF data were simulated by delaying the source signals (upsampled at 97.65 MHz) to the array elements according to their start time and one-way propagation path (equations 1-2), and convolving them with the array directivity term $h_m(\boldsymbol{r}_{\boldsymbol{S},l}, t)$. The latter was modelled using the array spatial impulse response from the built-in function in Field II [34], [35]. The simulated RF data were then downsampled to 9.76 MHz, and experimental RF recordings in the absence of vaporization events were used as additive noise. The average resulting Signal to Noise Ratio (SNR) was 16 dB. A total of 49 acquisitions were simulated, with one to seven sources per acquisition and seven acquisitions for each case.

## B. Experiments: passive imaging of ND vaporizations triggered by proton radiation

NDs with a decafluorobutane core and a polyvinyl alcohol shell (median diameter of 799 nm and polydispersity index of 0.3 [8]) were diluted in water to concentrations of 10 or 20 µM of PFB and irradiated with a clinical pulsed proton beam (Particle proton therapy center, Leuven university hospital, Belgium), as shown in Fig. 1. The proton pulses were delivered by a superconducting synchrocyclotron (S2C2, IBA, Louvain-la-Neuve, Belgium) with pulse length of ~10 µs and a pulse repetition frequency (PRF) of 1 kHz [36], [37]. Phantoms were irradiated with a monoenergetic proton Gaussian pencil beam at 70 MeV (2 phantoms) or 71 MeV (1 phantom). The nozzle current varied from 0.06 to 0.1 nA, corresponding to dose rates between 0.7 and 1 Gy/s in the Bragg peak.

The irradiated sample was inside a polymethyl methacrylate container of dimensions 85 × 70 × 80 $mm^3$ located above a magnetic stirrer set to 250 rpm to ensure constant renewal of NDs in the irradiated region, and a heated control plate to maintain the temperature to the desired value of 50°C. The water tank was aligned with the proton beam using in-room lasers and markings at the target beam position.

The 11 × 11 matrix array described in section III.A was attached to an acoustically-transparent window on the side of the water tank, and positioned such that the azimuth axis (US-x) was parallel to the proton beam with the proton range position located approximately at its center (Fig. 1). The ultrasound elevation and depth axis (US-y and US-z, respectively), were perpendicular to the proton beam. Passive RF data were acquired by an ULA-OP 256 research scanner (X-Phase srl, Florence, Italy) at a sampling rate of 19.5 MHz (acquisition length of 82 µs). The ULA-OP acquisitions were synchronized with the pulsed beam delivery using a trigger signal received from the Synchrocyclotron's Real-time and Sequencing Electronic Unit (SRSEU) at the proton PRF (1 kHz). Hence, every acquisition corresponded to the delivery of a single pulse of protons in the sample. The irradiation duration was controlled manually and lasted 1-4 seconds.

### C. Data analysis

All the data processing steps were implemented in Matlab (R2024a, The Mathworks, Natick, MA, USA). The RF data were preprocessed with a high-pass filter (1 MHz cutoff frequency, Butterworth order 6) prior to the localization algorithms to restrain the signals to the probe bandwidth (1.3-2.7 MHz). The M-step of SAGE was performed using Matlab's *fmincon* optimization routine. The conventional SAGE formulation (in Appendix) was used for simulated data, while experimental data suffered from a higher degree of residual interference and were processed with the SAGE variant of section II.C2. The used hyperparameters are detailed in Table 1. As benchmark comparison, the RF data were beamformed with 4D TD-DAS + CF (using (3) and (4)) and sources were localized by peak detection in 4D, followed by 3D intensity-weighted center of mass (IWCOM) super-localization [38] of the PSF center on the brightest temporal frame index.

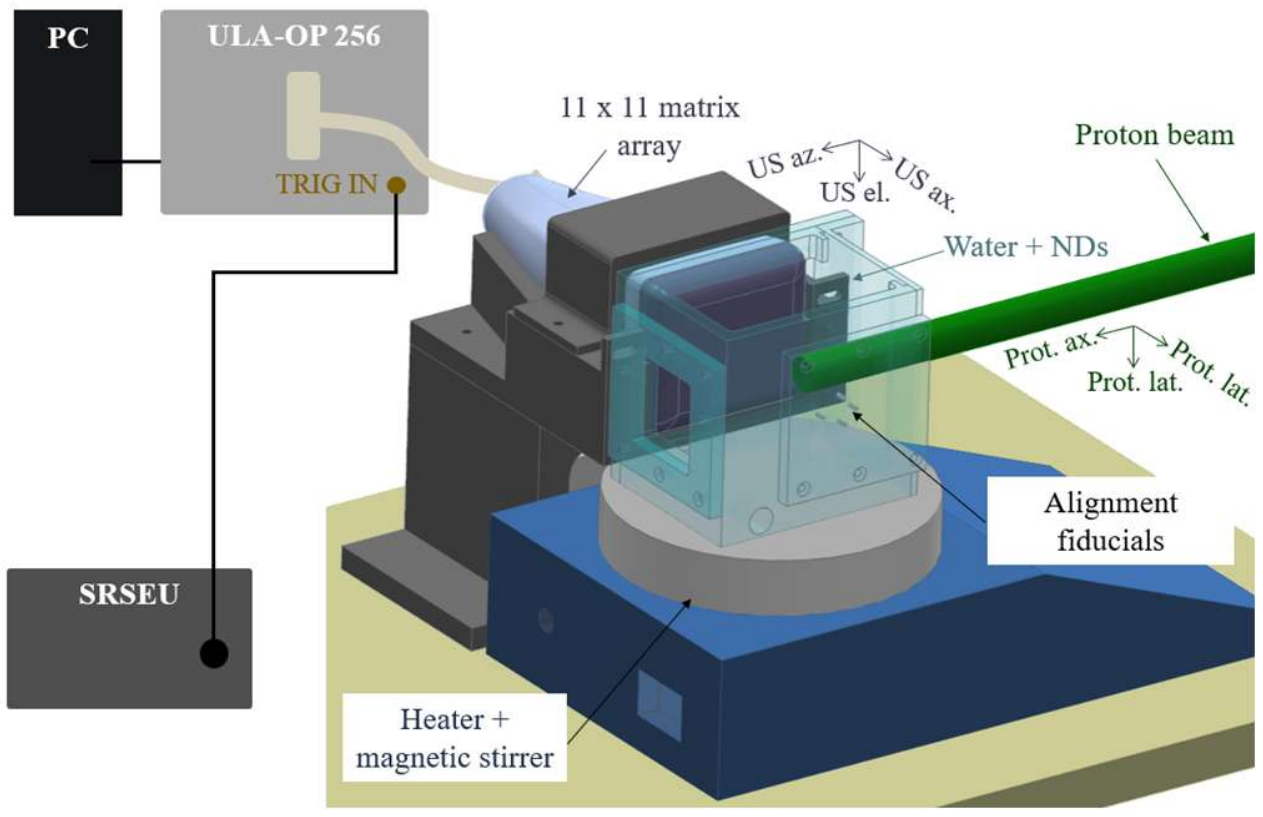

**Fig. 1:** Experimental setup used for passive imaging of ND vaporization events triggered by proton irradiation.

**Table I:** List of used parameters

| Parameter | Simulations | Experiments |
|---|---|---|
| Receive apodization | Tukey with F-number = 1.1 | |
| Beamformed volume $x \times y \times z$ [mm] | 30 × 30 ×50 | 32 × 34 ×50 |
| Maximum number of sources $L_{\max}$ | 7 | 5 |
| Neighbour distance thresholds $x \times y \times z$ [mm] | 2 × 2 × 4 | 2 × 2 × 6 |
| $t_0$ range for source initialization [µs] | 12 | -6 to +8 |
| Distance exit criterion $d_\theta$ [mm] | 1 | |
| SSE divergence threshold $d_{\mathrm{SSE}}$ | 1.5 | |
| Max number of SAGE iterations $k_{\max}$ | 10 | |

#### 1. *Simulations*

The source coordinates obtained via the three considered localization methods (4D TD-DAS+CF, GSE, SAGE) were paired to the true coordinates to measure the localization error. False negatives (missed events) were recorded when a source was not localized within an absolute distance of 15 mm, and false positives were recorded when an extra source was found with no corresponding ground truth. A Wilcoxon signed-rank test was used to determine whether one method yielded lower absolute localization errors than the other (one-sided, α=0.05). Differences in localization precision between methods were assessed using paired bootstrap resampling (10 000 iterations). The 95% confidence interval of the standard deviation difference was obtained from the empirical bootstrap distribution.

#### 2. *Experiments*

Experimental RF data were preprocessed with bandpass filtering (150 kHz – 3 MHz) followed by a spatiotemporal filter as described in [33], and downsampled to 9.76 MHz. After localization, vaporization event coordinates were registered to room coordinates (*i.e.* the reference coordinates for the proton beam) using translation and rotation factors derived from imaging four M2 set screw fiducials inserted in the phantom. Next, vaporization maps were compared to the 3D distribution of the proton stopping positions, as previously described in [33]. Briefly, analytical models were combined with measurements of the proton range using an ionization chamber in water (StingRay, IBA Dosimetry) and the lateral entrance spot size (Lynx PT, IBA Dosimetry). Maximum-likelihood estimation was used to determine the distances-to-agreement between the proton stopping distribution and the vaporization positions, as well as the Gaussian beam width ($\sigma$) of the vaporization distribution, along all three spatial directions. The vaporization yield (number of vaporizations normalized by the proton fluence and ND concentration) was computed based on the delivered proton fluence retrieved from the machine irradiation log files.

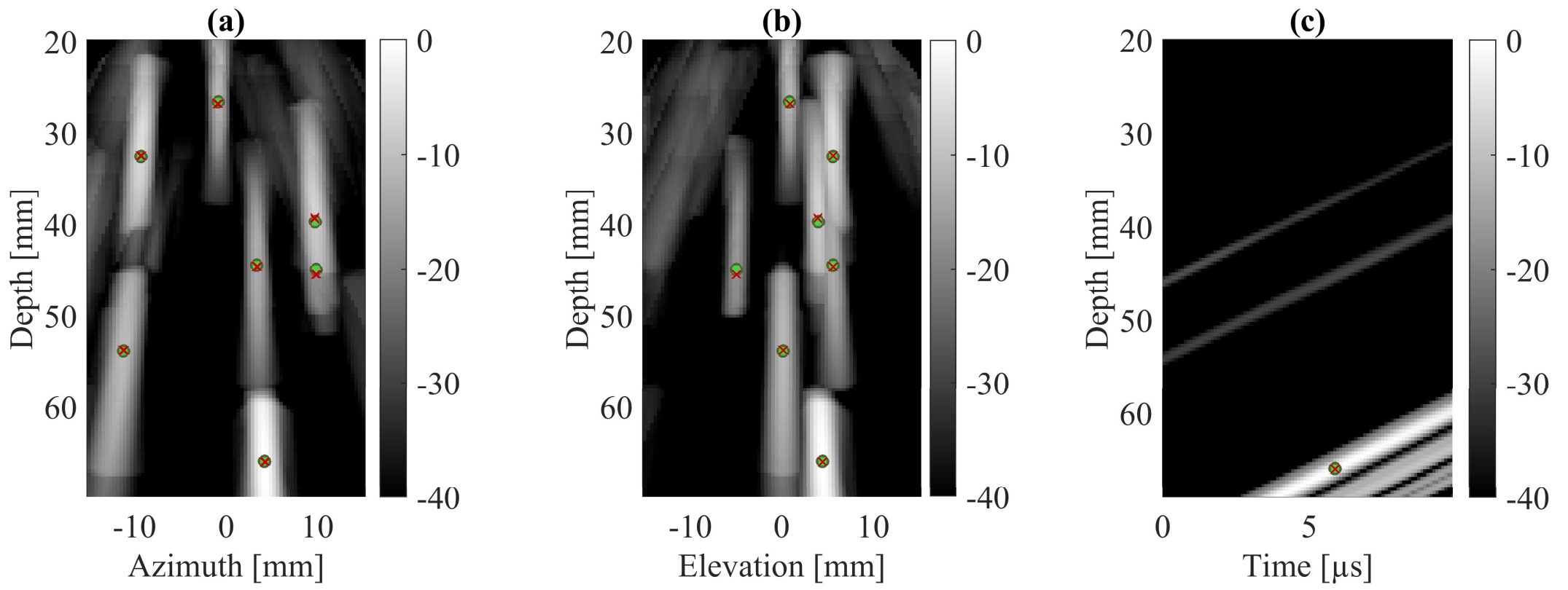

**Fig. 2:** Example of beamformed datacube (4D TD-DAS+CF) for a simulation with 7 sources, represented as maximum intensity projections on the *x*-*z* (a), and *y*-*z* (b) planes, and *z*-*t* plane cut at the strongest source (x,y) position (c) (grayscale in dB). Green circles represent the ground truth space and time ($t_0$) coordinates and red crosses are the parameter values obtained using SAGE.

## IV. Results

### A. Simulations

Fig. 2 shows the 4D beamformed data (TD-DAS+CF) for a simulation with 7 sources as 2D maximum intensity projections in *x*-*z* (a) and *y*-*z* (b). The beamforming time window was restricted to 12 µs (8 µs of $t_0$ range + 4 µs of potential timing uncertainties as would be the case in experiments), which leads to PSFs elongated in the axial direction. In Fig. 2c, which shows a 2D projection along the *z*-*t* axis for the strongest source, the source appears to be moving towards the probe as the time reference for beamforming progresses, as expected for TD-DAS applied at successive time points. The ground truth source location (and $t_0$) is shown by the green circle, while parameters estimated by SAGE are represented by red crosses.

Out of the 196 simulated acoustic sources, 93% were successfully detected by the GSE and SAGE methods, and only one source with no corresponding ground truth (false positive) was localized. The localization error as a function of the number of sources per acquisition is displayed in Fig. 3 for the azimuth (a), elevation (b) and axial (c) direction, for the three considered methods. The standard deviation of the localization error appears to be relatively independent from the number of sources per acquisition and was comparable in azimuth and elevation, while higher by an order of magnitude in the axial direction. The mean azimuth and elevation localization errors were <0.02 mm for all methods, and the mean axial localization error was -0.71 mm for the benchmark method (TD-DAS+CF followed by super-localization), 0.12 mm for the GSE method and 0.04 mm for the SAGE method. Fig. 3d shows boxplot distributions for the total localization error for the benchmark method, the GSE method and the SAGE method; their standard deviations were 1.22, 0.52, and 0.24 mm, respectively. Both the localization accuracy (median error) and precision (standard deviation of the error) were significantly improved by SAGE ($p<0.05$) compared to the benchmark and GSE methods.

### B. Experiments

Fig. 4 compares the spatial distribution (2D projection on the US *y*-*z* plane, transverse to the beam direction) of ND vaporization events localized during irradiation of a proton pencil beam, for the three localization methods: TD-DAS+CF beamforming followed by super-localization (a), GSE (b), and SAGE (c). For the benchmark approach (a), the strong directivity of the large-elements matrix array leads to preferential diagonal near-field 'streaks', absent for the other two methods, which use a propagation model accounting for array directivity. The SAGE map (c) bears the closest resemblance to the expected Gaussian shape of the proton beam, as illustrated in Fig. 5a-b which provides an overlay of the vaporization events localized by SAGE on the proton stopping distribution. The estimated ND emission time ($t_0$) distribution is shown in Fig. 5c and visually resembles the expected proton pulse shape (10 µs duration) [37].

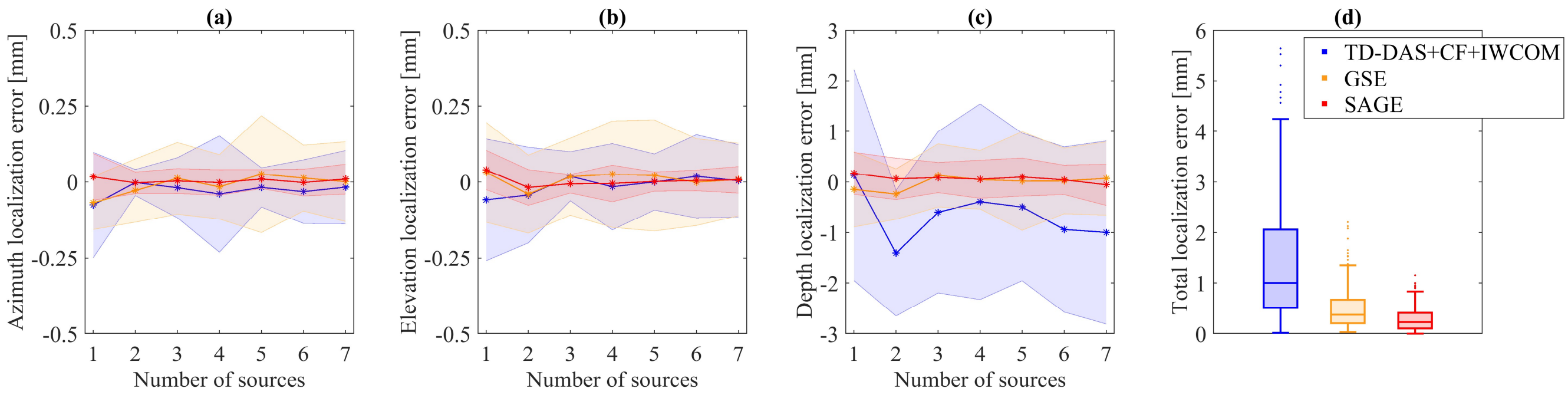

**Fig. 3:** Source localization error in the azimuth (a), elevation (b) and axial (c) direction as a function of the number of sources per acquisition. (d) Boxplots of the total localization errors grouped for all the simulations (benchmark method in blue, GSE in orange, SAGE in red).

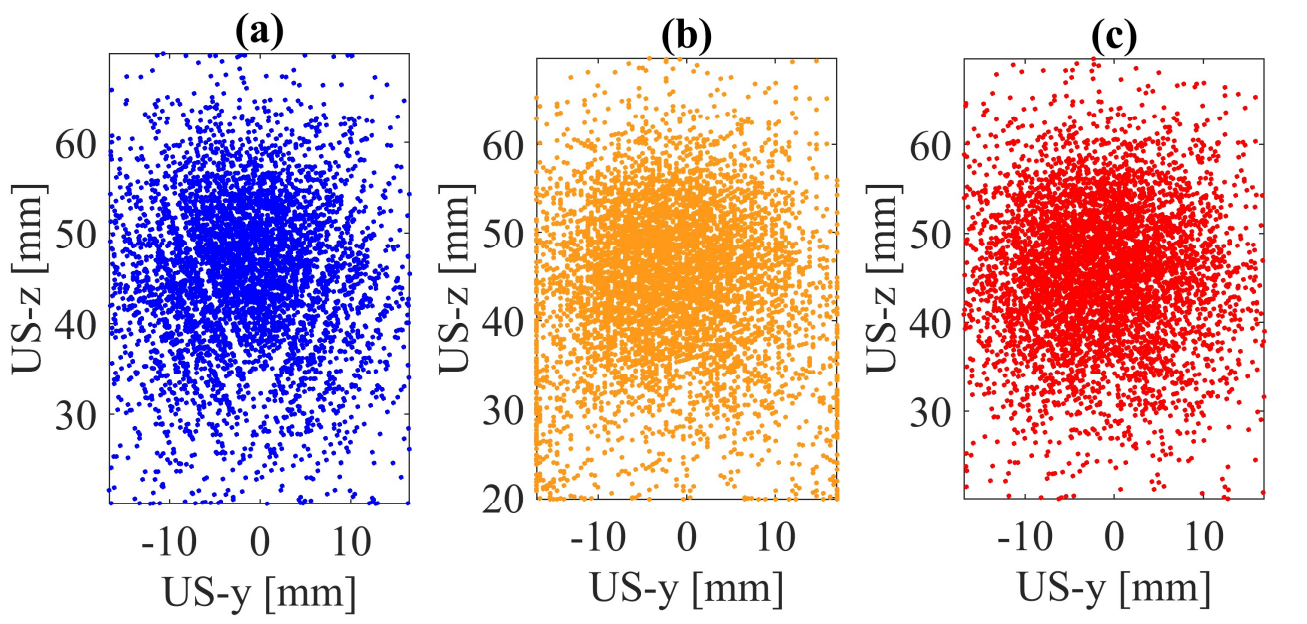

**Fig. 4:** 2D projections in the US *y-z* plane of ND vaporization events recorded during irradiation with a Gaussian proton pencil beam, and localized using (a) TD-DAS+CF followed by super-localization, (b) the GSE algorithm, and (c) the SAGE algorithm.

The SAGE method localized between 0 and 5 events per acquisition or proton pulse. The average vaporization yield is displayed in table II, and its evolution as irradiation progresses is shown in Fig. 6: a constant yield throughout irradiation is expected, as the vaporization count is proportional to the proton fluence [39], [40]. Table II also reports the experimental settings and localization results for each phantom. Localization accuracy was quantified by the distance-to-agreement between the estimated and reference proton range position (US-x) and proton spot center (US-y and US-z), as well as by the difference between the fitted and true Gaussian beam widths (σ). Distances-to-agreement were below 0.5 mm in all 3 directions. The fitted beam width differed from the true beam width by less than ±0.05 mm in the proton range direction (US-x) and by less than ±1 mm in the proton lateral directions (US-y and US-z). Fig. 7 displays the expected transverse (a-b) and axial (c) proton stopping distributions (green), together with binned vaporization positions and their fitted 1D profiles (full and dashed lines, respectively) for phantom 1 (red) and 2 (magenta). The phantoms were irradiated at different energies of 70 and 71 MeV, corresponding to an expected range shift of 1.05 mm; the range shift measured from the vaporization maps was 1.00 mm (Fig. 7c).

## V. Discussion

In this paper, we present a parametric method for ultrasound source localization in 3D + time based on the SAGE algorithm, suitable for spatially sparse sources without requirements on the acoustic emissions duration and absolute timing. SAGE localization performances were tested on simulated RF data with up to 7 sources in the field of view, and on experimental data of ND vaporization induced by proton radiation. The source signals (ND acoustic signatures) had variable center frequencies, durations, amplitudes, and emission times ($t_0$). Simulation results indicate that the standard deviation of the total SAGE localization error can be as low as 0.24 mm when the emission time uncertainty is 12 µs. In ATP-PAM or traditional DSAI imaging [15], such an uncertainty translates to a minimum axial resolution of 1.8 cm (further expanded by the emitted signal duration). Compared to previous studies in which ND vaporization events were temporally isolated, enabling their super-localization on RF data using TDoA fitting [33], [41], the proposed SAGE method extends this Passive-ULM concept to the localization of multiple temporally-overlapping sources with sub-PSF precision. Our SAGE parametric approach also exhibited robustness against strong inter-channel waveform distortions arising from the large element directivity and experimental conditions. In contrast to methods based on explicit estimation of wavefront arrival times, such as the hyperbolic fitting approach [42], [43], the present framework directly fits the complete channel data through a propagation model accounting for the array spatial impulse response. Consequently, localization does not rely on a separate arrival-time estimation step and may be less sensitive to distortions affecting waveform shape and timing consistency across channels.

In simulations, SAGE localization performances were significantly superior (5 fold improvement in the standard deviation of the total localization error) to 4D TD-DAS+CF beamforming followed by a super-localization approach (IWCOM), in particular in the axial direction (Fig. 3). Source localization by SAGE was also significantly improved (by a factor 2) compared to the GSE method, also used as source initialization step for SAGE. This was further evidenced by experimental data, which revealed that the benchmark method led to a near-field distortion of the expected 2D Gaussian proton spot by diagonal streaks (Fig. 4). We hypothesize that these streaks result from the interaction between the super-localization procedure and the diffraction-induced spatial structure of the beamformed PSF. Localization of each source is performed on the frame corresponding to the maximum beamformed intensity, making the estimated source positions sensitive to inhomogeneities in the beamformed field. Due to the large element pitch (3.6λ) and limited directivity of the matrix array, the beamformed intensity distribution is strongly influenced by the array diffraction pattern. Accounting for the array spatial impulse response in the system model suppressed this effect both for GSE and SAGE localizations (Fig. 4b-c). However, GSE alone produced erroneous localizations near the edges of the field of view, likely due to increased grating lobes, whereas these artifacts appeared to be correctly suppressed or reassigned during the subsequent SAGE optimization. Simulations also show the robustness of the localization error as the number of sources per acquisition increased (Fig. 3); however, the number of missed sources increased for higher number of sources per acquisition.

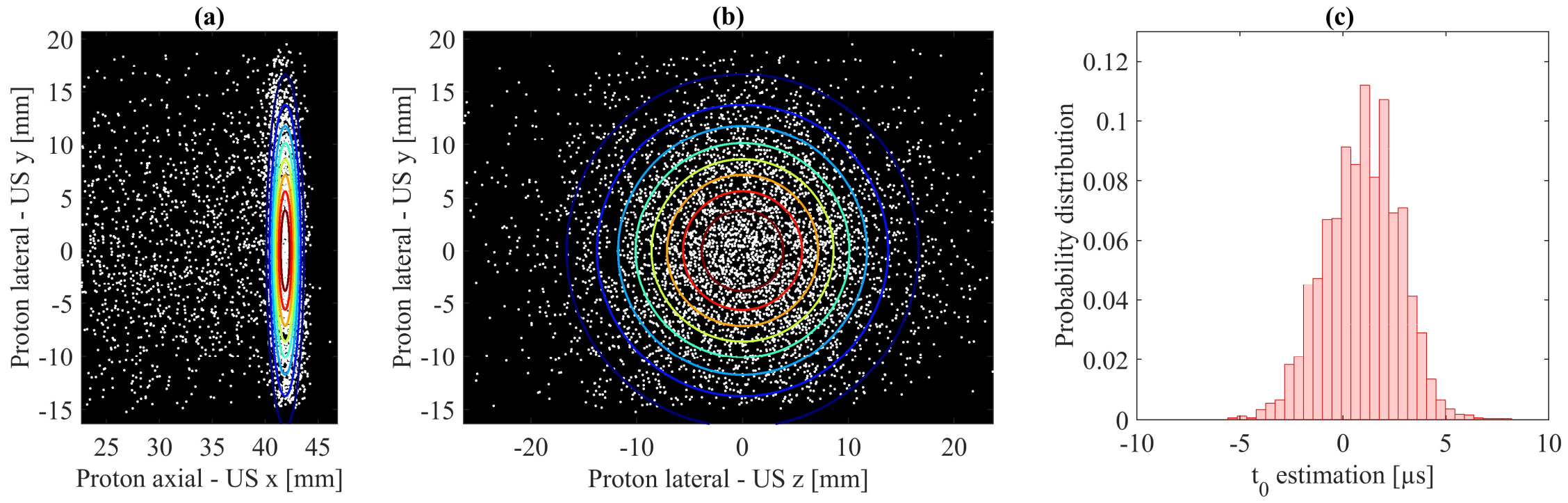

**Fig. 5:** (a-b) Individual ND vaporizations (white dots) localized using SAGE during the irradiation of a proton Gaussian pencil beam and overlaid on the proton stopping distribution (contour) projected onto the US *x-y* plane (a) and US *y-z* plane (b). The US-x (azimuth) direction is parallel to the proton beam direction, and the axes are provided in the beam reference system (the proton axial axis corresponds to the depth travelled by the protons in water, and the transverse axes origin is located at the center of the proton spot). (c) Distribution of the estimated emission time ($t_0$) of individual ND vaporizations triggered by pulsed proton irradiation and localized using the proposed SAGE algorithm. The distribution is expected to follow the proton pulse shape, which has a reported pulse length of ~10 µs [37]. The $t_0$ = 0 corresponds to the expected timing of the proton pulse based on a diode measurement.

**Table II:** Experimental results

| Phantom | | 1 | 2 | 3 |
|---|---|---|---|---|
| Beam Energy [MeV] | | 70 | 71 | 70 |
| ND concentration [μM] | | 20 | 20 | 10 |
| Number of delivered protons per pulse | | $6.3\times10^5$ | $6.4\times10^5$ | $4\times10^5$ |
| Number of processed acquisitions (= proton pulses) | | 2573 | 2573 | 4893 |
| Number of localized vaporization events | | 5247 | 4755 | 727 |
| Vaporization yield (events per $10^8$ protons per μM NDs) | | 12.5 | 11.8 | 2.7 |
| Distances to agreement [mm] | US-x (range difference) | 0.11 | 0.06 | -0.02 |
| | US-y | 0.46 | 0.07 | -0.12 |
| | US-z | -0.38 | -0.01 | 0.22 |
| Fitted beam width difference [mm] | US-x | 0.04 | 0.05 | -0.03 |
| | US-y | -0.57 | -0.24 | -0.03 |
| | US-z | 0.17 | 0.83 | 0.73 |

In experiments, SAGE enabled to accurately recover the spatiotemporal distribution of the delivered proton beam (Fig. 5), with a temporal resolution as high as the RF data sampling rate. In the absence of ground truth values for individual vaporizations, we used the vaporization distribution fit to the proton beam distribution to evaluate the localization accuracy. Distances-to-agreement or translational differences between the vaporization and proton distributions were all below 0.5 mm, which is within the experimental setup uncertainty budget (estimated to be ± 2 mm due to the visual setup alignment using in-room lasers and the registration using fiducial markers on the ultrasound image). In the US-x (proton range) direction, where the proton stopping distribution is the narrowest ($\sigma$ = 0.85 mm), the fitted beam width was within ±0.05 mm of its true value for all phantoms, highlighting the ability of SAGE to precisely localize vaporizations. This is further evidenced by the 0.05 mm range shift retrieval precision between phantom 1 and 2 (Fig. 7c). In the US-y and US-z directions, the proton distribution is wider ($\sigma$ = 8 mm at 70 MeV, Fig. 7a-b), which might explain the higher (but still sub-mm) differences in fitted beam width, as for the same number of counts, the fitting uncertainty increases with the Gaussian beam width. The larger error (up to 0.83 mm, Table II) in the US-z direction might be due to the inherent higher axial localization error of SAGE.

Several limitations must be acknowledged. Due to the relatively wide lateral proton distribution (in US-y and US-z), those experimental results represent a first proof-of-concept of the feasibility of the SAGE approach in a real-life scenario rather than a complete experimental validation of the SAGE localization accuracy. Therefore, the localization accuracy reported in simulations should be further experimentally validated with dedicated setups. In addition, experiments were performed in a water-based phantom, in which attenuation, scattering and speed of sound variations were only induced by background microbubbles present in the sample. Real tissue conditions might further distort the emitted signals and worsen the localization accuracy.

Of note, the benchmark method used a super-localization approach based on the IWCOM, which might not be the optimal approach for localizing passive acoustic emissions. Other ULM methods (onset [44], radial symmetry [45]), or super-localization methods developed specifically for passive imaging [5], might improve the axial localization performances of the benchmark method.

The validity of the presented SAGE parametric model is limited to sparse point sources and might not apply to dense cavitation clouds, for which alternative approaches might be preferred [24]. We expect our method to be best suited to applications involving transient sparse events *e.g.* inertial cavitation events [5] or phase-change events with a relatively low UCA concentration or probability of cavitation. While our results show stable performances for up to 7 simultaneously emitting sources per acquisition, higher number of sources might lead to increased localization errors or missed sources due to a stronger spatiotemporal overlap of their acoustic emissions, which complicates the source initialization step. The situation worsens for acoustic emissions of long durations or if the considered emission time window for beamforming increases. Indeed, long axial tails lead to localization performances degrading for sources located in the same propagation path from the source to the probe, as the initialization step might mix their emitted signals.

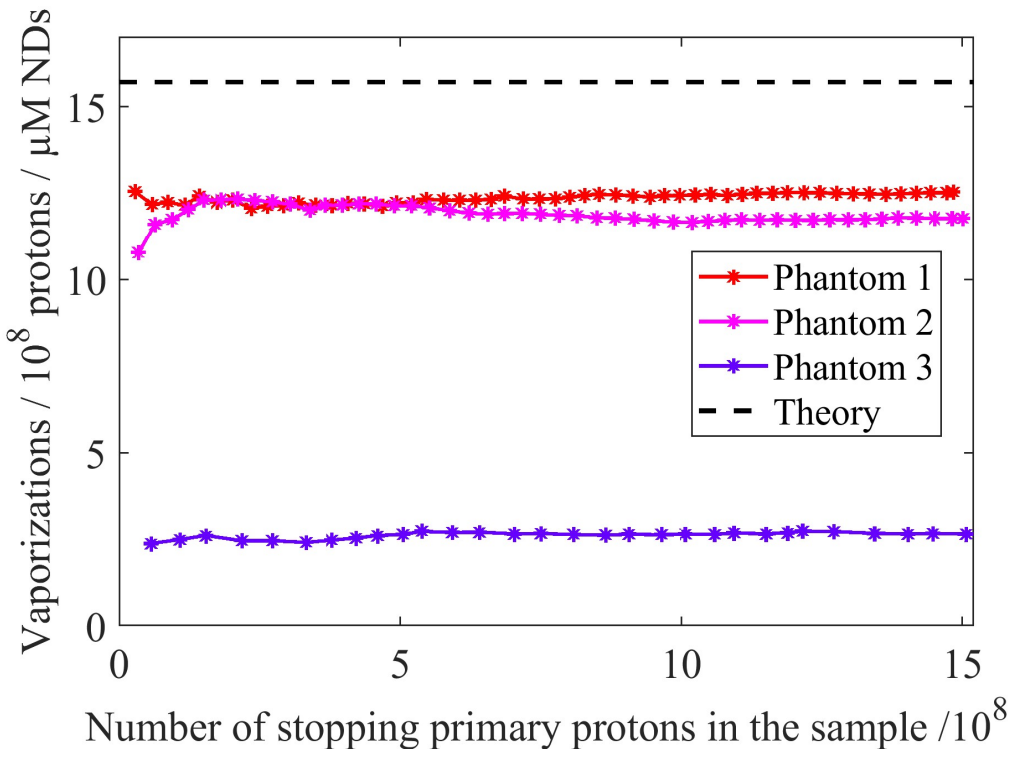

**Fig. 6:** Vaporization yield as a function of the number of primary protons stopping in the ultrasound field of view. The black dashed line is the vaporization yield expected from the theory of radiation-induced nucleation.

The neighbour joining step was introduced both as a mitigation to imperfect source cancellation and as a means to avoid this mixing of emitted signals, but implies that closely-located sources cannot be resolved, which may be problematic when close sources are expected. This neighbour joining step could potentially be removed by adopting a parametric model for the source emissions. Additionally, we hypothesize that the imperfect source cancellation was largely due to the high directivity of the array elements, leading to a strong signal distortion for elements located at large lateral distances from the sources. We partially countered that effect in the SAGE variant of Algorithm 2, but such mitigations might not be required for arrays with a pitch closer to λ or λ/2, for which a simultaneous optimization of $\boldsymbol{r}_{\boldsymbol{s},l}$ and $s_l(t)$ (as in Algorithm 3) might be more performant.

To obtain a tractable least-squares estimation framework in the M-step of the SAGE algorithm, a white Gaussian noise model was assumed [28]. In practice, after subtraction of the currently estimated source contributions, the residual signal generally contains electronic noise, modelling errors, and contributions from unmodelled sources, and might not be strictly white or Gaussian. Consequently, the maximum-likelihood interpretation of the least-squares estimator and its associated optimality properties might no longer strictly hold [46]. The localization accuracy may then be affected by increased estimator variance and, in the presence of structured modelling errors, by estimation bias. If the residual covariance matrix is known or can be estimated, the least-squares cost could be replaced by a generalized least-squares criterion weighted by the inverse covariance matrix, thereby potentially improving estimation efficiency [46].

The number of sources was not estimated a priori but chosen based on successive source subtraction with a threshold on the residual channel data energy and an user-defined upper bound $L_{\max}$. While several methods are available in the literature for estimating the number of sources [47], [48], they require access to second-order statistics (covariance matrix or CSM) through multiple realizations or averaging over long durations [26] or subapertures [49], which is not applicable to short, transient emissions and highly directional arrays. In simulations, the low percentage of missed sources (<10%) and near-absence of extra localizations confirms that our method can reliably estimate the number of sources. In experiments, we observed a vaporization yield constant throughout the irradiation (Fig. 6), in agreement with the theory [39] and indicating that the estimated number of sources was consistent. Compared to the theoretical yield (black dashed line in Fig. 6), vaporization rates were up to 30% lower in phantoms 1 and 2 and drastically lower in phantom 3. The latter might be explained by the fact that this specific phantom was irradiated at a temperature of 47°C instead of the 50°C target; we previously observed that the vaporization rate is strongly influenced by temperature [50]. The ~30% decrease in vaporization rate in other phantoms might be due to experimental ND concentration variations, and lack of detection of events with very low center frequencies due to the preprocessing high-pass filter applied on the RF data.

The presented SAGE approach, which we chose to implement in the TD, is also applicable in the FD. The TD choice was motivated by the relatively short duration (< 30 µs) and variability in frequency content of the ND vaporization signals, which complicates the selection of appropriate frequencies for each source, as they are simultaneously emitting. The TD implementation also provides the option to select an expected time window in the GSE algorithm, which decreases the localization uncertainty. In contrast, FD-SAGE might be more suitable when frequency selection is more straightforward (*e.g.* single source situations, emission frequency linked to excitation frequency), decreasing the number of source parameters to consider. Of note, the electromechanical impulse response of the receive chain is included in $s_l(t)$, which should be interpreted as the emitted pressure waveform convolved with the transducer transfer function. Deconvolution may be performed after source localization to recover an estimate of the emitted pressure waveform.

Future work will focus on improving computational efficiency to aim at real-time operation, and reducing the remaining assumptions of the source model, including the neighbour joining strategy and the white-noise approximation. We also plan to validate the proposed method in increasingly complex experimental conditions, including tissue-mimicking and *in vivo* environments. Beyond the presented application of radiotherapy monitoring, this SAGE framework may also prove valuable for other inertial cavitation monitoring or phase-change contrast agent imaging applications.

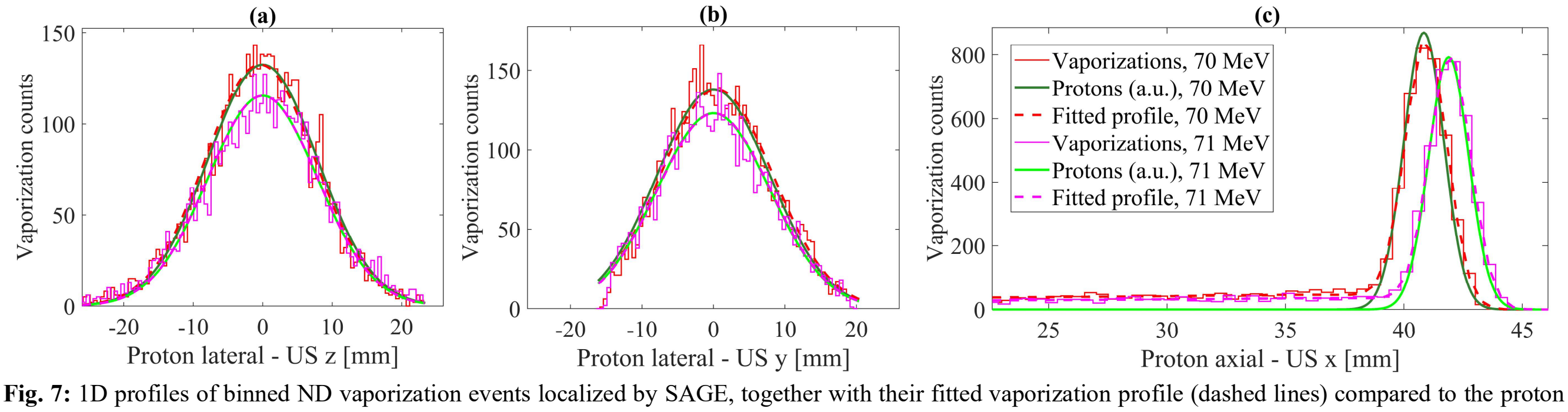

**Fig. 7:** 1D profiles of binned ND vaporization events localized by SAGE, together with their fitted vaporization profile (dashed lines) compared to the proton stopping distribution obtained from independent beam measurements and beam modelling (section III.c) (full green lines). The profiles are shown in the directions transverse (a-b) and parallel (c) to the proton range direction. Two phantoms irradiated at different energies of 70 MeV (red) and 71 MeV (magenta) are compared, leading to an expected shift of 1.05 mm in the proton range direction (c).

## VI. Conclusion

This paper presented a parametric method, based on the SAGE algorithm, to estimate the location and acoustic emissions of sparse cavitation nuclei. Our method is particularly suitable for the passive mapping of transient acoustic emissions, and exhibits robustness to uncertainties on the exact timing of these emissions. Owing to the source parametrization, the method achieves localization performances beyond the diffraction limit. In addition, the inclusion of the array spatial impulse response allows to correct for diffraction effects that might negatively affect super-localization applied to spatiotemporal maps beamformed with conventional DAS. By combining physically grounded source modeling with iterative parameter estimation, the proposed method opens new opportunities for diffraction-unlimited localization and quantitative characterization of transient acoustic emissions.

## VII. Appendix: SAGE algorithm for joint estimation of source position and waveform

The SAGE adaptation proposed in section II.C2 decouples the estimation of source position and waveform to improve robustness against residual interference arising from imperfect source cancellation. For completeness, we also present here a conventional SAGE formulation in which all source parameters are updated jointly in the M-step (Algorithm 3).

Starting from the M-step described in (11), we substitute a closed-form solution of $s_l$ as a function of $\boldsymbol{r}$, which allows to minimize only for $\boldsymbol{r}$ (*i.e.* for 3 parameters instead of $N_t$+3). Indeed, (11) is linear in $s_l$ and the objective function can be written in the frequency domain as:

$$\widehat{\boldsymbol{\theta}}_l^{(k)} = \arg\min_{\boldsymbol{\theta}} \sum_{m=1}^{M} \sum_{f=1}^{N_f} \left( \hat{Y}_{m,l}^{(k)}(f) - G_m(\boldsymbol{r}, f) S_l(f) \right)^2, \quad (17)$$

With $\hat{Y}_{m,l}^{(k)}(f)$ and $S_l(f)$ the Fourier transforms of $\hat{y}_{m,l}^{(k)}(t)$ and $s_l(t)$, and $G_m(\boldsymbol{r}, f)$ the transfer function made of the product of the Fourier transform of the spatial impulse response $H_m(\boldsymbol{r}, f)$ and the propagation delay term $e^{-j2\pi f \;\, m(\boldsymbol{r})}$. For fixed $\boldsymbol{r}$, the least-squares solution for the source signal is:

$$\hat{S}_l(\boldsymbol{r}, f) = \frac{\sum_m G_m^*(\boldsymbol{r}, f) Y_{m,l}(f)}{\sum_m |G_m(\boldsymbol{r}, f)|^2 + \beta \frac{P_N(f)}{P_S(f)}} \quad (18)$$

which corresponds to a matched-filter estimate followed by an inversion of the receive transfer function. A Wiener regularization term (with β an empirical scaling factor and $P_N(f)$ and $P_S(f)$ the noise and signal power spectral densities) is included to stabilize this inversion at frequencies at which $|G_m(\boldsymbol{r}, f)|$ is small, thereby limiting the contribution of poorly observable frequencies that would otherwise be amplified by the inversion. This effect is particularly beneficial for arrays with large, highly directional receive elements, for which the spatial impulse response may strongly attenuate high-frequency components.

The time-domain source signal estimate is obtained via inverse Fourier transform of (18). Substituting this estimate into (11) reduces the optimization to the three spatial coordinates. After updating the parameters $\widehat{\boldsymbol{\theta}}_l^{(k)}$, the estimated channel data for source $l$ is updated via the system model:

$$\hat{y}_{m,l}^{(k)}(t) = h_m\left(\hat{\boldsymbol{r}}_{S,l}^{(k)}, t\right) * \hat{s}_l^{(k)}\left(t - \tau_m(\hat{\boldsymbol{r}}_{S,l}^{(k)})\right) \quad (19)$$

The algorithm exits if the parameter distance $d_\theta$ is below a threshold or the maximum number of allowed iterations is reached.

**Algorithm 3 – SAGE for joint estimation of source position and waveform**

**Require:** $y_m(t), h_m(\boldsymbol{r}, t), L, \boldsymbol{\Theta}^{(0)}, \hat{y}_{m,l}^{(0)}(t), k_{\max}$, parameter threshold $d_\theta$

Initialize: $k = 1$

**While** $k \leq k_{\max}$ **do**

  **For** $l = 1, \dots, L$ **do**

    $\hat{y}_{m,l}^{(k)}(t)$ via (10) **E-step**

    $\hat{\boldsymbol{r}}_{S,l}^{(k)}$ via (11), with $\hat{s}_l(\boldsymbol{r}, t)$ from ifft(18) **M-step**

    $\hat{s}_l^{(k)}(t)$ from ifft(18)

    $\hat{y}_{m,l}^{(k)}(t)$ via (19)

  **End for**

  Neighbour merging

  **If** $\max_l \|\hat{\boldsymbol{r}}_{S,l}^{(k)} - \hat{\boldsymbol{r}}_{S,l}^{(k-1)}\| < d_\theta$ **do**

    **exit**

  **else**

    $k = k + 1$

  **End if**

**End while**

**Return** $L, \boldsymbol{\Theta}^{(k)}, \hat{y}_{m,l}^{(k)}(t)$

## VIII. Acknowledgment

The authors would like to thank Dr. Bram Carlier, Dr. Marcus Ingram and Dr. Kenneth Poels for their assistance with the experimental data acquisition, Prof. Tom Depuydt and Prof. Edmond Sterpin for providing access to the clinical proton therapy facilities. They would also like to thank the Leuven IBA team and Mr. Nicolas Gerard for the synchronization with the proton beam, and Profs. Jan D'hooge and Koen Van Den Abeele for stimulating discussions and support.